\documentclass{wiley-article}
\usepackage[numbers]{natbib}
\usepackage{siunitx, amsmath}
\usepackage{algorithm, algpseudocode}
\usepackage{natbib}
\usepackage{float}
\usepackage{placeins}
\usepackage{multirow}
\usepackage{hyperref}
\setcitestyle{authoryear,open={(},close={)}}

\title{Poisson Count Time Series}

\author[1]{Jiajie Kong}
\author[2]{Robert Lund}

\contrib[\authfn{1}]{Equally contributing authors.}

\affil[1]{Department of Statistics, University of California, Santa Cruz, Santa Cruz, CA, 95064, United States}
\affil[2]{Department of Statistics, University of California, Santa Cruz, Santa Cruz, CA, 95064, United States}

\corraddress{Jiajie Kong, Department of Statistics, University of California, Santa Cruz, Santa Cruz, CA, 95064, United States}
\corremail{jkong7@ucsc.edu}

\presentadd[\authfn{2}]{Department of Statistics, University of California, Santa Cruz, Santa Cruz, CA, 95064, United States}

\fundinginfo{The authors thank NSF support from Grant DMS 2113592.}

\runningauthor{Kong}

\begin{document}

\begin{frontmatter}
\maketitle
 
\begin{abstract}
This paper reviews and compares popular methods, some old and some very recent, that produce time series having Poisson marginal distributions.  The paper begins by narrating ways where time series with Poisson marginal distributions can be produced. Modeling nonstationary series with covariates motivates consideration of methods where the Poisson parameter depends on time. Here, estimation methods are developed for some of the more flexible methods.
The results are used in the analysis of 1) a count sequence of tropical cyclones occurring in the North Atlantic Basin since 1970, and 2) the number of no-hitter games pitched in major league baseball since 1893.  Tests for whether the Poisson marginal distribution is appropriate are included.

% Please include a maximum of seven keywords
\keywords{Copulas, Count Time Series, Integer Autoregression, Poisson Distribution, Superposition, Thinning}
\end{abstract}
\end{frontmatter}

\section{Introduction}
Gaussian time series have a long and storied development in time series modeling \citep{box2015time, Brockwell_Davis_1991, shumway2000time}.  Indeed, most time series connoisseurs regard Gaussian theory as essentially complete now.  Less developed, but now currently heavily researched, are methods that describe autocorrelated series for counts; that is, the series $X_t$ at time $t$ is supported on some subset of the non-negative integers $\{ 0, 1, \ldots \}$. This paper reviews, compares, and contrasts several popular methods that produce Poisson distributed series, which is arguably the quintessential count distribution.  Discrete and integer autoregressions, superpositioning methods, and copula methods are considered.

Some caveats are worth mentioning at the onset.   First, techniques exist that produce count models having a conditional Poisson distribution. One such technique, which is essentially the GLARMA paradigm of \cite{davis2005maximum, dunsmuir2015glarma}, starts with a nonnegative $\{ \lambda_t \}$ process that is stationary and posits that the conditional distribution of $X_t |\lambda_t$ is Poisson with mean $\lambda_t$. While this can often lead to a convenient autoregressive representation of the counts \citep{fokianos2009poisson}, $X_t$ will not be left with a Poisson marginal distribution.  Indeed, should $X|\Lambda=\lambda$ be Poisson distributed with mean $\lambda$, then the marginal distribution of this structure must be overdispersed:
\[
\mbox{Var}(X)= E[ \mbox{Var}(X|\Lambda)]+\mbox{Var}(E[X|\Lambda])= E[\Lambda] + \mbox{Var}(\Lambda) > E[\Lambda]=E[X].
\]
We refer the reader to the review in \cite{davis2021count} and the references within for more on this issue. In particular, this paper focuses on models having a true Poisson count marginal distribution.

As a second caveat, some results for stationary Gaussian time series do not hold in the Poisson setting.  For one example, if $\{ \gamma(h) \}_{h=-\infty}^\infty$ is a symmetric $(\gamma(h) = \gamma(-h)$ for all integers $h \geq 0$) and non-negative definite sequence on the integers, then there exists a Gaussian distributed sequence $\{ X_t \}$ with $\mbox{Cov}(X_t, X_{t+h})=\gamma(h)$.  No such result carries over to the Poisson case.  Indeed, $(-1)^h$ is symmetric and non-negative definite.   While a Gaussian sequence with this autocovariance exists (take $X_t=(-1)^tZ$, where $Z$ is standard normal), it is not possible to achieve this in the Poisson setting.  To see this, it is enough to show that one cannot have two Poisson variables $X_1$ and $X_2$ having the same mean $\lambda$ and correlation $-1$ (the reader is challenged to prove this).

The rest of this paper proceeds as follows. The next section reviews methods that generate count series having a Poisson marginal distribution.  There, discrete and integer autoregressions, superpositioning methods, and copula techniques are considered.  The pros and cons of each model classes are illuminated; much of this material constitutes a review.  Section \ref{Estimation} moves to estimation issues.  There, likelihood estimation techniques are developed if possible.   Unfortunately, the joint distribution needed in the likelihood is intractable for many model classes.  Particle filtering and quasi-likelihood techniques such as linear prediction will be used here.  Simulations show that the methods work quite well.  Section \ref{App} analyzes series of annual North Atlantic Basin hurricanes and no-hitter games pitched in major league baseball with covariates. Section \ref{Diagnostics} shows how to test whether the Poisson marginal distribution assumption is adequate with residual diagnostics.  Section \ref{Comments} concludes the paper with comments. 

\section{Methods}
\label{Methods}
This section reviews methods producing a stationary series having Poisson marginal distributions.  Some of this material has appeared elsewhere; however, some new insights are offered in our discourse.  

As some of the models classes below cannot have negative autocorrelations, flexibility and completeness of the autocovariances becomes an issue. Before proceeding, we first investigate the most negatively correlated Poisson variables existing, providing some intuition en route.

Let $F_\lambda(\cdot)$ be the Poisson cumulative distribution function (CDF) with mean $\lambda$:
\[
F_{\lambda}(n) = \sum_{k=0}^n \frac{e^{-\lambda} \lambda^k}{k!}, \quad n = 0, 1, \ldots 
\]
The most negatively correlated pair of random variables $X$ and $Y$, both having the marginal cumulative distribution function (CDF) $F_{\lambda}$, are known to have form $X=F_{\lambda}^{-1}(U)$ and $Y=F_{\lambda}^{-1}(1-U)$, where $U$ is uniformly distributed over [0,1] and $F_{\lambda}^{-1}$ is the inverse CDF:
\[
F_{\lambda}^{-1}(u)= \inf \{ t : F_{\lambda}(t) \geq u \}
\]
\citep{whitt1976bivariate} (this version of the inverse is the quantile function). Such an $(X,Y)$ pair can be produced from a Gaussian copula via
\[
X=F_{\lambda}^{-1}(\Phi(Z)), \quad 
Y=F_{\lambda}^{-1}(\Phi(-Z)).
\]
Here, $\Phi(\cdot)$ is the standard normal CDF and $Z$ is a standard normal random variable.  This is because $U:=\Phi(Z)$ is uniformly distributed over [0,1] by the probability transformation theorem and $\Phi(-Z) = 1 -\Phi(Z)=1-U$.

\begin{figure}[h!]
\centering
\includegraphics[width=7cm]{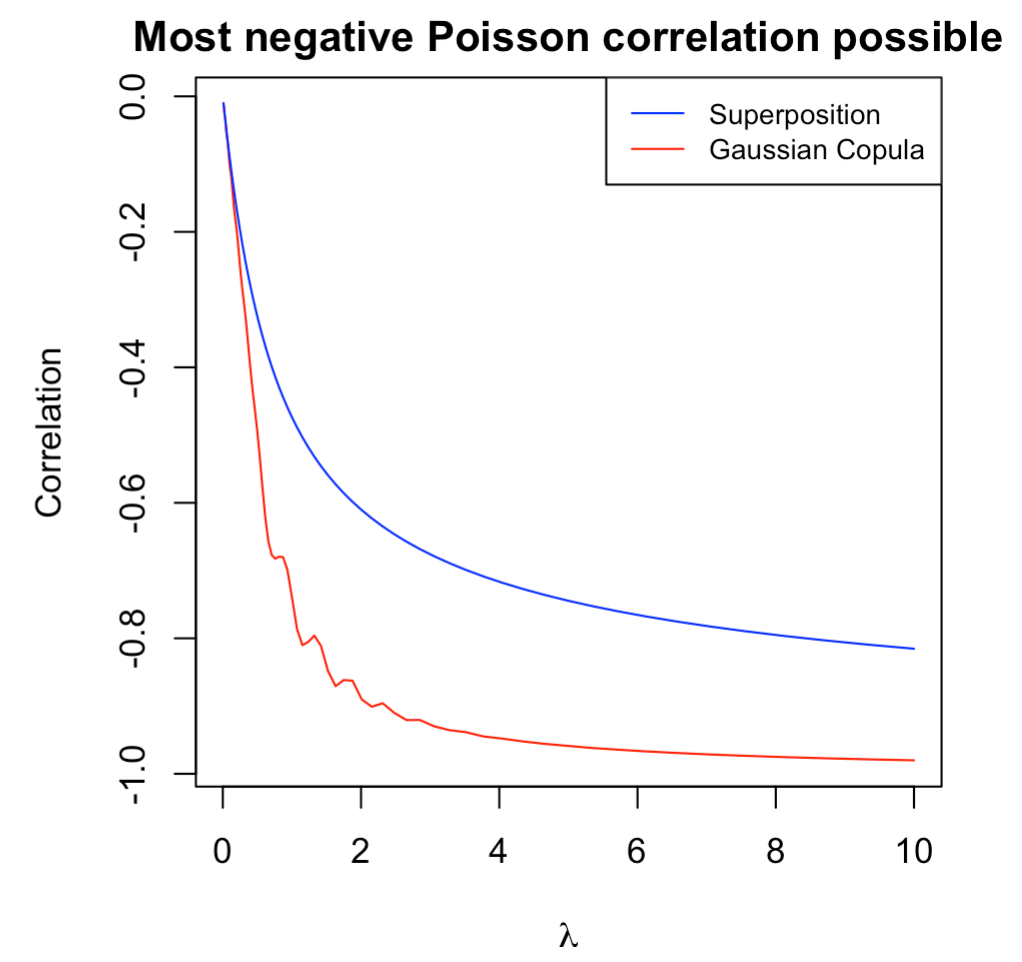}
\caption{Lower Curve:  A plot of the most negative achievable autocorrelation, $\mbox{NB}(\lambda)$, for various $\lambda$.   Bottom Curve: A plot of the most negative achievable autocorrelation in the superpositioned model class of Section \ref{super}.}
\label{fig:NB_lambda_plot}
\end{figure}

To obtain an expression for the most negative correlation possible, let $c_n=F_\lambda(n)$ denote the Poisson $\lambda$ CDF at index $n$ and note that the inverse has form
\[
F_{\lambda}^{-1}(u) = \sum_{n=1}^\infty 
n 1_{[c_{n-1}, c_n)}(u),
\]
where $1_A(x)$ denotes an indicator function over the set $A$.  Converting this to a tail sum gives
\begin{equation}
\label{hold1}
F_{\lambda}^{-1}(u) 
= \sum_{n=1}^\infty \sum_{k=1}^n 1_{[c_{n-1}, c_n)}(u)
= \sum_{k=1}^\infty \sum_{n=k}^\infty 1_{[c_{n-1}, c_n)}(u)
= \sum_{k=1}^\infty 1_{[c_{k-1}, 1)}(u).
\end{equation}
Simple algebraic manipulations now give
\[
E[ F_{\lambda}^{-1}(U) F_{\lambda}^{-1}(1-U) ]=
\sum_{k=1}^\infty \sum_{\ell=1}^\infty 
E[1_{[c_{k-1},1)}(U) 1_{[c_{\ell-1},1)}(1-U) ]=
\sum_{k=0}^\infty \sum_{\ell=0}^\infty (1-c_{\ell}-c_{k}) 1_{ [c_{\ell} + c_{k} < 1]}.
\]
An expression for the most negative autocorrelation, which we denote by $\mbox{NB}(\lambda)$, now follows simply as 
\begin{equation}
\label{NegativeBound}
\mbox{NB}(\lambda)=\frac{ \sum_{k=0}^\infty \sum_{\ell=0}^\infty (1-c_{\ell}-c_{k}) 1_{ [c_{\ell} + c_{k} < 1]} - \lambda^2}{\lambda}.
\end{equation}
A plot of $\mbox{NB}(\lambda)$ as a function of $\lambda$ is provided in Figure \ref{fig:NB_lambda_plot}.  As $\lambda \rightarrow \infty$, this correlation tends to -1; however, for small $\lambda$, there are significant restrictions on the negative correlations that can be made.  An interesting feature of Figure \ref{fig:NB_lambda_plot} lies with the slight non-monotonicity of $\mbox{NB}(\lambda)$ in $\lambda$ for some $\lambda \leq 3$. This is not computational roundoff; indeed, the Hermite coefficients $g_k$ below, discussed in Subsection 2.4, are not monotonic in $\lambda$.  This fact can also be inferred from the plots in the supplementary material in \cite{jia2021count}.

\subsection{Discrete Autoregressions}

Discrete autoregressions (DARs), the original attempt to devise stationary series having a particular marginal distribution \citep{jacobs1978discrete_a, jacobs1978discrete_b, jacobs1978discrete_c}, work by mixing past series values.  In the Poisson case, the construction begins with a sequence $\{ A_t \}_{t=1}^\infty$ of IID Poisson variables with mean $\lambda > 0$.  A sequence $\{ B_t \}_{t=1}^\infty$ of IID Bernoulli trials is needed that is independent of $\{ A_t \}$ and has success probability $P(B_t=1) \equiv p$.

In the first-order case, the DAR construction starts by taking $X_1=A_1$. For $t \geq 2$, series values are mixed via
\[
X_t = B_t X_{t-1} + (1-B_t) A_t.
\]
Here, if $B_t=1$, $X_t$ is taken as $X_{t-1}$; should $B_t=0$, $X_t=A_t$ is a ``new independent Poisson draw".  Schemes extending the paradigm to higher autoregressive (AR) orders are achievable with additional Bernoulli sequences, but the DAR class has some drawbacks. Foremost, DAR models cannot have any negative autocorrelations.  In the first order case, one can show that $\mbox{Corr}(X_t, X_{t+h})= p^h$ for $h \geq 0$, which cannot be negative since $p \in (0,1)$ must be a probability.  Perhaps worse, series values are often repeated:  $P[X_{t+1}=X_t] \geq p$.  In the heavily correlated case where $p$ is close to unity, the series becomes almost constant.  Because of these properties, DAR series were essentially abandoned. See \cite{moller2020generalized} for recent attempts to remedy these issues.  We will not consider the DAR class further.

\subsection{Integer Autoregressions}

\cite{steutel1979discrete} introduced binomial thinning in an attempt to mimic AR recursions for count series. If $X$ is a count-valued variable, define $\alpha \circ X = \sum_{i=1}^X B_i$, where $\{ B_i \}$ are IID Bernoulli trials with success probability $\alpha \in (0,1)$ that are independent of $X$; $\circ$ is called a binomial thinning operator.

Integer autoregressions (INARs) are based on thinning operators.  In the first-order case, a strictly stationary series with Poisson marginal distributions with mean $\lambda$ is governed by the difference equation
\begin{equation}
\label{eq:INAR(1)}
X_t = \alpha \circ X_{t-1} + \epsilon_t, 
\end{equation}
where $\{ \epsilon_t \}$ is IID with a Poisson marginal distribution with mean $\lambda(1-\alpha)$ \citep{mckenzie1985some, alzaid1990integer, weiss2018introduction}. 

A well-known property of solutions to (\ref{eq:INAR(1)}) is that any discrete self-decomposable marginal distribution (these include Poisson, negative binomial, and generalized Poisson) can be produced by this recursion; this said, our focus remains on Poisson marginals. The autocorrelation function of an INAR(1) series can be shown to have the form
\begin{equation}
\mbox{Corr}\left( X_t, X_{t+h} \right) = \alpha^h;
\end{equation}
in particular, negative autocorrelations cannot be produced since $\alpha \in (0,1)$.

Higher order schemes, dubbed INAR($r$) for order $r$, have been investigated; however, producing series with Poisson marginals in these schemes has been problematic.  Be wary of issues with the literature here; specifically, the methods in \cite{alzaid1990integer} and \cite{du1991integer} will not achieve Poisson marginals; see the discussion in \cite{scotto2015thinning}.  To circumvent this problem, \cite{zhu2006modelling} propose combined integer autoregressive (CINAR) models.  A CINAR series of order $r$ follows the recursion
\begin{equation}
\label{eq:CINAR(p)}
X_t =   D_{t,1} \left( \alpha \circ X_{t-1} \right) + \dots + 
        D_{t,r} \left( \alpha \circ X_{t-r} \right) + \epsilon_t.
\end{equation}
The IID time $t$ ``decision vector" is $\boldsymbol{D}_t = (D_{t,1}, \dots, D_{t,r}) \sim \mbox{Mult}(1; \phi_1, \dots, \phi_r)$ and is independent of $\{ \epsilon_t \}$ and $\{ X_s \}_{s<t}$. The decision vector $\boldsymbol{D}_t$ chooses which of the past $r$ series values is used in the thinning, enabling the scheme to keep a Poisson marginal distribution.  Here, the innovation $\epsilon_t$ and the thinning of $X_{t-j}$ for the chosen $j \in \{ 1, \ldots, r \}$ are conducted independently. 

\cite{zhu2003new} show that the marginal distribution of any CINAR($r$) series must also be self-decomposable.  As with the Poisson INAR(1) model in (\ref{eq:INAR(1)}), the CINAR($r$) model has a marginal Poisson distribution with mean $\lambda$ when $\{ \epsilon_t \}$ is IID Poisson with mean $\lambda (1-\alpha)$, regardless of the order of $r$. \cite{weiss2008combined} derives the autocovariance of a CINAR($r$) series from (\ref{eq:CINAR(p)}); from this, one can show that the resulting autocovariance must be non-negative.  As such, CINAR($r$) models cannot have any negative autocorrelations and this model class also fails to span the range of all possible autocovariances.

\subsection{Superposition Techniques}
\label{super}

Poisson distributions can be built by adding IID copies of Bernoulli trials.  Indeed, if $\{ B_i \}_{i=1}^\infty$ are IID Bernoulli variables with success probability $p=P[B_t=1]$ and $N$ is Poisson, independent of $\{ B_i \}_{i=1}^\infty$ and with mean $\lambda$, then $\sum_{i=1}^N B_i$ has a Poisson distribution with mean $p \lambda$.  \cite{blight1989time} and \cite{cui2009new} use this construction to produce correlated count series having Poisson marginal distributions.

Elaborating, suppose that $\{ B_{t,i} \}_{t=1}^\infty$ are IID copies of the autocorrelated Bernoulli trial sequence $\{ B_t \}$ for $i \geq 1$. Clarifying, for each fixed $i$, $\{ B_{t,i} \}_{t=1}^\infty$ is autocorrelated in time $t$ --- say $\mbox{Cov}(B_{t,i}, B_{t+h,i}):= \gamma_B(h)$ --- but $\{ B_{t,i} \}_{t=1}^\infty$ and $\{ B_{t,j} \}_{t=1}^\infty$ are independent when $i \ne j$.  A series with Poisson marginals can be built via superpositioning:
\[
X_t = \sum_{i=1}^{N_t} B_{t,i}.
\]
Then $X_t$ has a Poisson distribution with mean $\lambda p$. The autocovariances of $\{ X_t \}$ are 
\begin{equation}
\label{ACFX}
\gamma_X(h):=\mbox{Cov}(X_t, X_{t+h})= 
E[ \min(N_t, N_{t+h}) ] \gamma_B(h)
\end{equation}
for $h > 0$, where $N_t$ and $N_{t+h}$ are {\bf independent} Poisson variables with mean $\lambda$ ($E[\min(N_t, N_{t+h})]$ does not depend on $h$). Note that $\gamma_X(h)$ will be negative whenever $\gamma_B(h)$ is negative; hence, this model class can produce negatively correlated series.  One can show that
\begin{equation}
\label{defC}
\kappa (\lambda) := E[ \min(N_t,N_{t+h})]=
2 \lambda \left[ 1-e^{-4 \lambda} 
\{ I_0(4 \lambda) + I_1(4 \lambda) \} \right],
\end{equation}
where the $I_j$s are modified Bessel functions of the first kind:
\[
I_j(x)=\sum_{n=0}^\infty \frac{(x/2)^{2n+j}}{ n!(n+j)!}, \quad j=0,1
\]
\citep{jia2021superpositioned}.  The above construct essentially builds the correlated Poisson series $\{ X_t \}$ from the independent Poisson series $\{ N_t \}$.

Several ways to construct correlated sequences of Bernoulli trials exist.  One way uses a stationary renewal sequence built from the IID lifetimes $\{ L_i \}_{i=1}^\infty$ supported on $\{ 1, 2, \ldots \}$ and an initial delay $L_0$ supported on $\{ 0, 1, 2, \ldots \}$ as follows.  Define the random walk $S_n = L_0 + L_1 + \cdots + L_n$ for $n \geq 0$ and set $B_t=1$ when a renewal occurs at time $t$ (i.e., when $S_n=t$ for some $n \geq 0$) and zero otherwise.  When the initial delay $L_0$ is chosen as the first derived distribution of one of the $L_i$s for $i \geq 1$ (a generic copy of these is denoted by $L$), viz.
\[
P[L_0=k]=\frac{P[ L > k ]}{\mu_L}, \quad k=0, 1, 2, \ldots, 
\]
the Bernoulli sequence is stationary in that $E[B_t] \equiv \tfrac{1}{\mu_L}$ and
\begin{equation}
\label{ACVF1}
\gamma_B(h) =
\frac{1}{\mu_L} \left( u_h - \frac{1}{\mu_L} \right).  
\end{equation}
Here, the notation has $\mu_L=E[L]$ and $u_h$ as the probability of a time $h$ renewal in a non-delayed renewal process ($L_0=0$).  When $L$ is aperiodic and $E[L] < \infty$ (which we henceforth assume), $u_h \longrightarrow \mu_L^{-1}$ as $t \longrightarrow \infty$ by the elementary renewal theorem \citep{smith1958renewal}.  One can show that $\{ X_t \}$ has long memory (absolutely non-summable autocovariances over all lags) when $E[L^2] = \infty$ \citep{lund2015long}; \cite{jia2021superpositioned} derive further properties of superpositioned series and investigate non-Poisson count marginal distributions.

Another way to produce a stationary but correlated Bernoulli sequence $\{ B_t \}$ clips a Gaussian sequence in the manner of \cite{kedem1980estimation}.  Elaborating, let $\{ Z_t \}$ be a standard stationary Gaussian sequence with $E[ Z_t ] \equiv 0$, $\mbox{Var}(Z_t) \equiv 1$, and $\mbox{Corr}(Z_t, Z_{t+h})=\rho_Z(h)$. Define Bernoulli trials via
\[
B_t = 1_A(Z_t),
\]
where $A$ is some fixed set.  In this case, autocovariances are, for $h \geq 0$,
\begin{equation}
\label{ACVF2}
\gamma_B(h) = P(Z_t \in A \cap Z_{t+h} \in A) - P(Z_t \in A)^2.
\end{equation}
As an example, when $A=(0, \infty)$, $P(Z_t \in A) = 1/2$ and classic bivariate normal orthant probability calculations give
\[
\gamma_B(h) = \frac{\arcsin(\rho_Z(h))}{2 \pi}.
\]
Notice that the autocovariances in (\ref{ACVF1}) and (\ref{ACVF2}) can be negative.  Specifically, for a renewal $\{ B_t \}$, $\gamma_B(h) < 0$ whenever $u_h < \mu_L^{-1}$; for a Gaussian clipped $\{ B_t \}$ with $A=(0, \infty)$, $\gamma_B(h) < 0$ whenever $\rho_Z(h) < 0$.

While the autocovariance function in (\ref{ACFX}) can be decisively negative, it does not achieve the minimum possible in (\ref{NegativeBound}). To see this, note from (\ref{ACFX}) and (\ref{defC}) that
\[
\rho_X(h) := \mbox{Corr}(X_t, X_{t+h}) = \frac{\kappa(\lambda/p)}{\lambda} \gamma_B(h).
\]
The smallest $\gamma_B(h)$ that can be made from a binary distibuted pair of random variables each having success probability $p$ can be shown to be -$p^2$ for any lag $h$. Thus, the most negative lag $h$ correlation that can be built from this model for any $h$ is $-p^2 \kappa(\lambda/p)/\lambda$.

Figure \ref{fig:NB_lambda_plot} displays this most negative correlation.  Again, the correlation approaches $-1$ as $\lambda$ increases and there are significant restrictions for $\lambda$ close to zero. To numerically calculate these minimums, a grid search was used to find the $p \in (0,1)$ that minimizes $-p^2 \kappa(\lambda/p)/\lambda$ for each fixed $\lambda$. These most negative correlations are uniformly bigger than the optimal ones identified earlier.  This brings us to our best Poisson model, which will achieve the full spectrum of achievable autocorrelations.

\subsection{Gaussian Copulas}

A recent class of very parsimonious and general count models developed in \cite{jia2021count} has been demonstrated to have remarkable flexibility.   This model starts with a stationary standard Gaussian sequence $\{ Z_t \}$ and transforms it to the desired count structure.  Standardized means that $E[ Z_t ] \equiv 0$, $\mbox{Var}(Z_t) \equiv 1$, and $\gamma_Z(h) = \rho_Z(h) = \mbox{Corr}(Z_t, Z_{t+h})$.  The construction transforms $Z_t$ at time $t$ via
\begin{equation}
\label{Trans1}
X_t = F_\lambda^{-1}(\Phi(Z_t)).
\end{equation}
Here, $\Phi(\cdot)$ is the standard normal CDF. By the probability integral transformation theorem, $\Phi(Z)$ has a uniform distribution over [0,1] and $X_t$ has a Poisson marginal distribution with mean $\lambda$.

The autocovariance function of $\{ X_t \}$ is unwieldy in form, but can be quantified through several expansions. With $G(x)=F_\lambda^{-1}(\Phi(x))$, arguing as in (\ref{hold1}) gives
\[
G(z)=\sum_{n=1}^\infty n 1_{[F_\lambda(n-1) \leq \Phi(z) < F_\lambda(n)]}
=\sum_{\ell=0}^\infty 1_{[F_\lambda(\ell-1),1)}(\Phi(z))= \sum_{\ell=0}^\infty 1_{[\Phi^{-1}(F_\lambda(\ell)), \infty)}(z).
\]
One expression for the autocovariances now follows as 
\[
\gamma_X(h)=\sum_{j=0}^\infty \sum_{k=0}^\infty 
P \left( Z_t > \Phi^{-1}(F_\lambda(j)) \cap
Z_{t+h} > \Phi^{-1}(F_\lambda(k)) \right),
\]
which is computationally intensive to evaluate.   

\begin{figure}[h!]
\centering
\includegraphics[width=7cm]{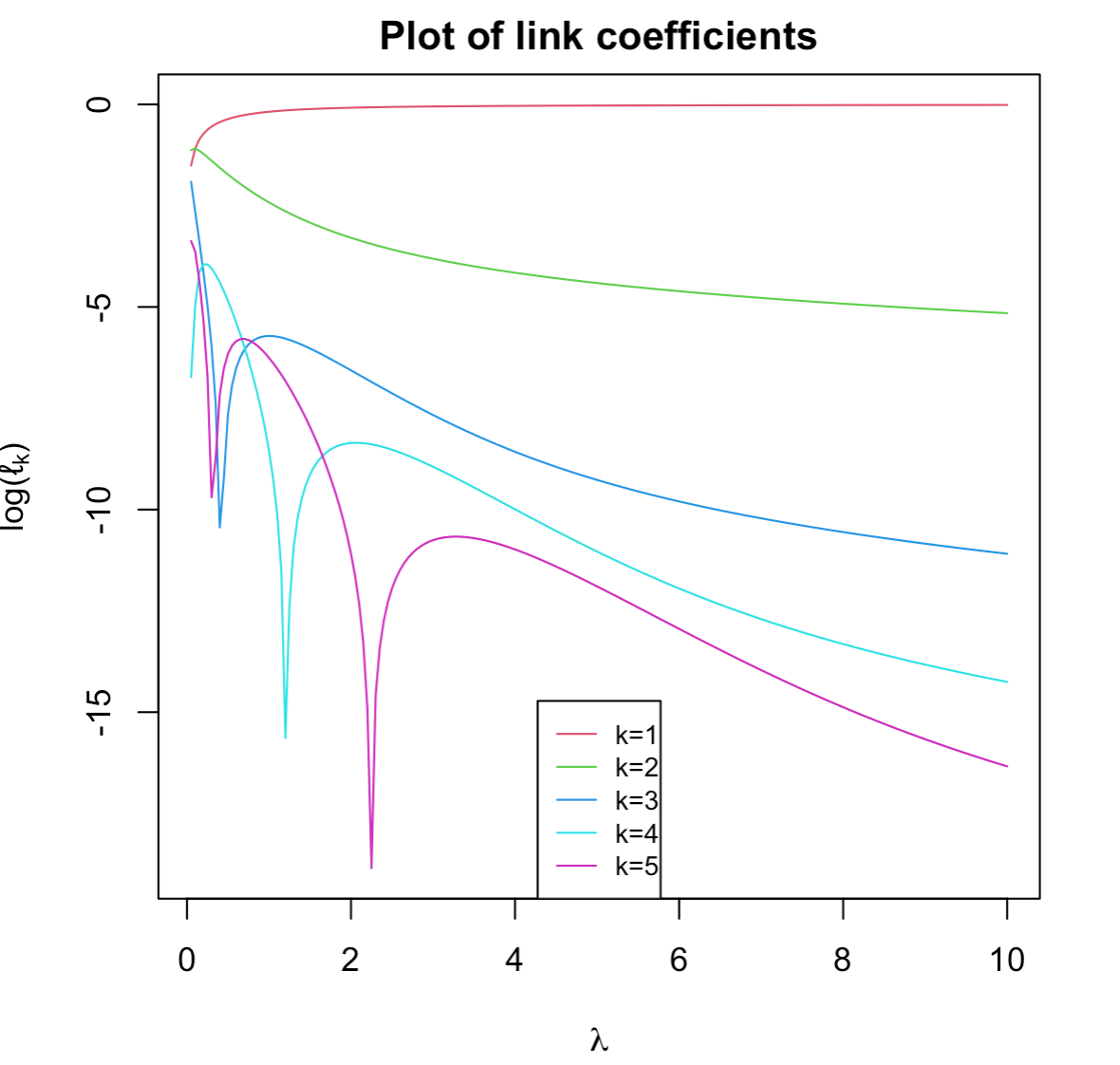}
\caption{Plot of $\log(\ell_k)$ versus $\lambda$ for $k \in \{1,2,3,4,5 \}$.}
\label{fig:link_coefficient_plot}
\end{figure}

A more tractable expansion works through the Hermite polynomials $\{ H_k(x) \}_{k=0}^\infty$ defined by
\[
H_k(x) = (-1)^ke^{z^2/2}\dfrac{d^k}{dz^k}\left( e^{-z^2/x} \right).
\]
The first three Hermite polynomials are $H_0(x) \equiv 1$, $H_1(x)=x$, and $H_2(x)=x^2-1$.  Higher order polynomials obey the recursion
\[
H_k(x) = xH_{k-1}(x) - H_{k-1}^\prime(x), \quad k \geq 1.
\]
Expanding $G$ in a Hermite basis, viz.
\[
G(x)= \sum_{\ell=0}^\infty g_\ell H_k(x),
\]
where $g_\ell$ is the $\ell$th Hermite coefficient
\[
g_{\ell}= \frac{E[G(Z) H_\ell(Z)]}{\ell!},
\]
and $Z$ is standard normal produces the key functional relationship between autocovariances in $\{ Z_t \}$ and $\{ X_t \}$:
\[
\gamma_X(h)=L(\rho_Z(h)).
\]
The function $L(\cdot)$ is called a link function in \cite{jia2021count} and has the power series representation
\[
L(u)=
\sum_{k=1}^{\infty}\dfrac{k!g_k^2}{\gamma_X(0)}u^k=
\sum_{k=1}^{\infty}\eta_ku^k, \quad |u| \leq 1,
\]
where $\eta_k=k!g_k^2/\gamma_X(0)$.  It is known that $L(u)$ is differentiable in $u$, that $L(0)=0$, and that $L(1)=1$ \citep{jia2021count}. The quantity $\eta_k$ is called a link coefficient. Figure \ref{fig:link_coefficient_plot} plots $\eta_k$ for a few values of $k$ as a function of $\lambda$.   While these coefficients ``behave erratically", they decrease quickly as $k$ and/or $\lambda$ increase. The value $\lim_{u \downarrow -1}L(u)$ is the most negative pairwise correlation that can be made.  The inequality $|\gamma_Z(h)| \leq |\gamma_X(h)|$, established in \citep{jia2021count} (see also \cite{tong2014probability}), shows that correlation will always be lost in the transformation from $Z_t$ to $X_t$; however, many times, this loss is not substantial. 

\begin{figure}[h!]
\centering
\includegraphics[width=7cm]{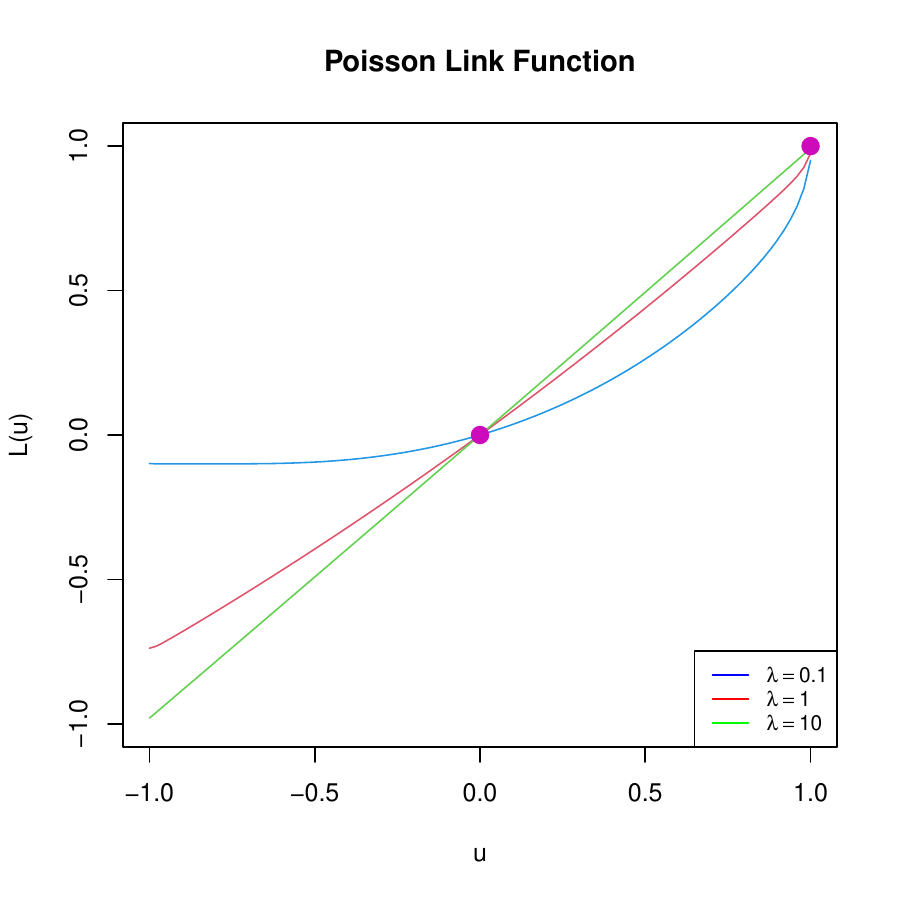}
\caption{The link function $L(u)$ for $\lambda \in \{0.1, 1, 10 \}$.}
\label{fig:link_function_plot}
\end{figure}

With the conventions $\Phi^{-1}(0)=-\infty$ and $\Phi^{-1}(1)=\infty$, the Hermite coefficients can be computed as
\[
g_k = \frac{1}{k!} \left[ 1- e^{-\lambda} +\sum_{\ell=1}^\infty
H_{\ell-1}(\Phi^{-1}(F_\lambda(\ell))) \phi(F_\lambda(\ell))
\right].
\]
Other forms for $g_k$ are derived in  \cite{jia2021count}. Figure \ref{fig:link_function_plot} plots $L(u)$ against $u$ for various values of $\lambda$. Notice that when $\lambda=10$, $L(u) \approx u$ and very little autocorrelation is lost in the transformation of $Z_t$ to $X_t$.

\section{Inference}
\label{Estimation}

In the stationary case where $X_t \sim \mbox{Poisson}(\lambda)$, the traditional estimate of $\lambda$ is the sample mean
\[
\hat{\lambda}= \frac{X_1 + \dots + X_n}{n}.
\]
The variance of this estimate is
\[
\mbox{Var}(\hat{\lambda})=
\frac{\lambda}{n} 
\left[ 1 + 2 \sum_{j=1}^{n-1} (1-j/n) \gamma_X(j) \right],
\]
which is consistent whenever $\{ X_t \}$ has short memory autocorrelations ($\sum_{j=1}^\infty |\rho_X(j)| < \infty)$.  One can get estimators of $\lambda$ with smaller variances than the sample mean via generalized least squares methods, but any improvements are negligible as $n \rightarrow \infty$ (see the discussion in \cite{chipman1979efficiency, lee2004revisiting}).

In practical modeling scenarios, $\{ X_t \}$ is usually non-stationary, possibly due to trends, periodicities, covariates, etc..  To study estimation for Poisson count series with such structures, time-varying versions of our techniques are needed. Hence, our immediate goal is to develop time-varying models where $X_t \sim \mbox{Poisson}(\lambda_t)$ and there is correlation between observations.

To develop such models, we first revisit INAR(1) models.  Here, complications immediately arise. To see this, if $X_1 \sim \mbox{Poisson}(\lambda_1)$ and the process obeys (\ref{eq:INAR(1)}), then we must have $\epsilon_2 \sim \mbox{Poisson}(\lambda_2-p\lambda_1)$.  Unfortunately, there is no guarantee that $\lambda_2-p\lambda_1$ is non-negative, suggesting that the INAR(1) paradigm is a suboptimal way to handle time-varying dynamics.   Should it be known that $\lambda_t$ is nondecreasing in $t$, then one could explore this model class further; see \cite{bentarzi2023periodic} for additional comments on process existence.  Because of this issue, we move to other methods.

\subsection{Time-varying Superpositioned Series}

In the superpositioned model class, time-varying models having the desired marginal properties are easy to construct.  For this, let $\{ N_t \}$ be a sequence of independent Poisson variables with $N_t \sim \mbox{Poisson}(\lambda_t/p)$. Here, $p = P[ B_{t,i}=1 ]$ is the success probability of the Bernoulli trials in the construction.  Then it is easy to see that $X_t \sim \mbox{Poisson}(\lambda_t)$ as required. In this case, the derivation associated with \ref{ACFX} gives, for $h > 0$,
\[
\mbox{Cov}(X_t, X_{t+h}) = 
\mathbb{E}(\min(N_t, N_{t+h}))\gamma_B(h),
\]
where $N_t$ and $N_{t+h}$ are independent Poisson variables with parameters $\lambda_t$ and $\lambda_{t+h}$, respectively. Note that
\begin{eqnarray*}
\mathbb{E}(\min(N_t, N_{t+h}))&=&\sum_{n = 1}^{\infty}P\left( \min(N_t, N_{t+h})\geq n \right)\\
&=& \sum_{n = 1}^{\infty}P\left( N_t \geq n \cap N_{t+h} \geq n\right)\\
&=&\sum_{n = 1}^{\infty}
P\left( N_t     \geq n \right) 
P\left( N_{t+h} \geq n \right)\\
&=&\sum_{n = 1}^{\infty}
[1-F_{\lambda_{t}/p}(n-1) ]
[1-F_{\lambda_{t+h}/p}(n-1)],
\end{eqnarray*}
where $F_{\lambda_t}(n-1)$ and $F_{\lambda_{t+h}}(n-1)$ are the respective Poisson CDFs at the times $t$ and $t+h$.   There does not seem to be a simplification of this formula as in (\ref{defC}) unless $\lambda_t=\lambda_{t+h}$. 

To estimate parameters in superpositioned schemes, we will use linear prediction methods.  Unfortunately, the model's likelihood function and conditional expectations $E[ X_t | X_1, \dots, X_{t-1}]$ appear to be intractable.  Also, how to simulate the likelihood accurately, as we will do for the Gaussian copula case below with particle filtering methods, is also unclear. A bivariate composite likelihood is tractable as an alternative to linear prediction; however, we will see that linear prediction works reasonably well.

Linear prediction works by first calculating $\mbox{Cov}(X_t, X_s):=\Gamma_X(t,s)$ for each $1 \leq t,s, \leq n$.  Estimators are found by minimizing the simple sum of squares
\begin{equation}
\label{eq:SSS}
\sum_{t=1}^n (X_t - \hat{X}_t)^2,
\end{equation}
where
\[
\hat{X}_t = \lambda_t + 
\sum_{j=1}^{t-1} w_{j,t} (X_j - \lambda_j)
\]
is the best one-step-ahead predictor of $X_t$ made from linear combinations of a constant and $X_1, \ldots, X_{t-1}$.  The prediction coefficients $\{ w_{j,t} \}$ satisfy the prediction equations
\begin{equation}
\label{YW}
\begin{bmatrix}
    \Gamma_X(1, 1)&\Gamma_X(1, 2)&\cdots&\Gamma_X(1, t-2)&\Gamma_X(1, t-1)\\
    \Gamma_X(2, 1)&\Gamma_X(2, 2)&\cdots&\Gamma_X(2, t-2)&\Gamma_X(2, t-1)\\
    \vdots&\vdots&\ddots&\vdots&\vdots\\
    \Gamma_X(t-2, 1)&\Gamma_X(t-2, 2)&\cdots&\Gamma_X(t-2, t-2)&\Gamma_X(t-2, t-1)\\
    \Gamma_X(t-1, 1)&\Gamma_X(t-1, 2)&\cdots&\Gamma_X(t-1, t-2)&\Gamma_X(t-1, t-1)
\end{bmatrix}
\begin{bmatrix}
    w_{1,t}\\
    w_{2,t}\\
    \vdots\\
    w_{t-2,t}\\
    w_{t-1,t}
\end{bmatrix}=\begin{bmatrix}
    \Gamma_X(1, t)\\
    \Gamma_X(2, t)\\
    \vdots\\
    \Gamma_X(n-2, t)\\
    \Gamma_X(n-1, t)
\end{bmatrix}.
\end{equation}

To obtain parameter estimators, the sum of squares is numerically minimized in the parameters appearing in $\{ \lambda_t \}$. In this scheme, we only minimize the sum of squares in (\ref{eq:SSS}) about the mean parameters appearing in $\lambda_t$; the parameters appearing in the covariance structure of $\{ B_t \}$ are held to their true values during this optimization.  Future work might consider how to estimate these parameters in tandem; here, a Cochrane-Orcutt recursion seems developable \citep{CO_1949} (the Gaussian copula structure analyzed next makes the issue somewhat moot). One complication is that some parameters in $\{ B_t \}$ (namely $p$) arise in both the mean and autocovariance structure of the linear predictors.  One may wish to consider weighted least squares to accommodate the changing variances of the series.

Solving the linear system in (\ref{YW}) requires a $\mathcal{O}(t^3)$ computational cost; as such, the computational burden can be expensive for large $n$. We recommend doing a Cholesky decomposition of the covariance matrix on the left hand side of (\ref{YW}) and then using backwards/forward substitution to obtain $\{ w_{t,k} \}$. The classic Durbin-Levinson recursion is not suitable here since $\{ X_t \}$ is not stationary. In our future computations, the ``Nelder–Mead" optimization method was used to minimize the sum of squares in (\ref{eq:SSS}).

\subsection{Time-varying Gaussian Copula Series}

In the Gaussian copula case, process construction carries through as before; specifically, we set
\begin{equation}
\label{Trans2}
X_t = F_{\lambda_t}^{-1} (\Phi(Z_t)).
\end{equation}
The Hermite expansion of the time homogeneous case is simply allowed to vary with time now.

For notation, let $\boldsymbol{\theta}$ contain all parameters appearing in $\{ \lambda_t \}_{t=1}^n$ and $\boldsymbol{\eta}$ denote all parameters governing $\{ Z_t \}$.  We do not suggest trying to incorporate time dependence into the dynamics of $\{ Z_t \}$ as process existence issues then arise.   The covariance matrix of $(Z_1, \ldots , Z_n)^\prime$ depends only on $\boldsymbol{\eta}$ (and not on $\boldsymbol{\theta}$).

The model's likelihood function, denoted by $\mathcal{L}(\boldsymbol{\theta}, \boldsymbol{\eta})$, is simply a high dimensional multivariate normal probability. To see this, use (\ref{Trans2}) with the data $X_1, \ldots , X_n$ to get
\begin{equation}
\label{eq:likelihood}
\boldsymbol{\mathcal{L}}( \boldsymbol{\theta},\boldsymbol{\eta})
= \mathbb{P}(X_1=x_1, \cdots, X_n = x_n)
= \mathbb{P}\left( Z_1 \in (a_1,b_1], \cdots, Z_n \in (a_n,b_n] \right),
\end{equation}
where $\{ a_t \}_{t=1}^n$ and $\{ b_t \}_{t=1}^n$ are
\[
a_t = \Phi^{-1}(F_{\lambda_t}(x_t-1)), 
\quad 
b_t = \Phi^{-1}(F_{\lambda_t}(x_t)).
\]

This probability is infeasible to accurately evaluate for large $n$.  A likelihood can, however, be quite accurately simulated by particle filtering methods \citep{douc2014nonlinear}.  Indeed, particle filtering simulation methods can be used to reliably approximate the model's likelihood and even compute standard errors. The current preferred methods of multivariate normal probability evaluation are arguably the Geweke–Hajivassiliou–Keane (GHK) simulators of \cite{geweke1991efficient} and \cite{hajivassiliou1996simulation}. Here, we develop an adaptive version of this simulator.

Particle filtering methods, which are classic importance sampling techniques, aim to evaluate integrals by drawing samples from an alternative distribution and averaging their corresponding weights. Should we need to estimate the integral $\int_D f(\boldsymbol{x})d\boldsymbol{x}$ over some domain $D$, then we use
\[
\int_D f(\boldsymbol{x})d\boldsymbol{x} =  
\int_D \dfrac{f(\boldsymbol{x})}
{q(\boldsymbol{x})}q(\boldsymbol{x})d\boldsymbol{x},
\]
where $f(\boldsymbol{x})/q(\boldsymbol{x})$ is the weight and $q$ is called the importance distribution. The importance sampling estimate of the integral is
\[
\widehat{\int_{D}\dfrac{f(\boldsymbol{x})}{q(\boldsymbol{x})}q(\boldsymbol{x})d\boldsymbol{x}} = \dfrac{1}{m}\sum_{k=1}^{m}\dfrac{f(\boldsymbol{x}^{(k)})}{q(\boldsymbol{x}^{(k)})},
\]
where $\boldsymbol{x}^{(1)}, \ldots, \boldsymbol{x}^{(m)}$ are $m$ IID samples drawn from $q$. We require that $q$ satisfies $q(z_{1:n}) > 0$ for $z_t \in (a_t, b_t]$ and $q(z_{1:n})=0$ otherwise; our notation uses $z_{1:k}=(z_1, \dots, z_k)$ and $x_{1:k}=(x_1, \dots, x_k)$.

We take advantage of the Markov chain properties of the latent AR $\{ Z_t \}$. The GHK algorithm samples $Z_t$, depending on the its previous history $Z_{t-1}, \ldots, Z_1$ and $X_t$, from a truncated normal density.  Specifically, let $p_{\boldsymbol{\eta}(t)} \left( z_t | z_{1:t-1}; x_{t} \right)$ denote the truncated normal density of $Z_t$ given the history $Z_1=z_1, \ldots, Z_{t-1}=z_{t-1}$ and $X_t=x_t$. Then
\begin{equation}
\label{eq:p_density}
p_{\boldsymbol{\eta}(t)} \left( z_t | z_{t-1}, \ldots, z_{1}, x_{t} \right) = \dfrac{1}{r_t}
\left[ 
\dfrac{\phi(\frac{z_t-\hat{z}_t}{r_t})}{\Phi(\frac{b_t-\hat{z}_t}{r_t})-\Phi(\frac{a_t-\hat{z}_t}{r_t})}
\right], \quad a_t < z_t < b_t, 
\end{equation}
where $\hat{z}_t$ and $r_t$ are the one-step-ahead mean and standard deviation of $Z_t$ conditioned on $Z_{1:t-1}$. Note that $a_t$ and $b_t$ only depend on $x_t$. We choose the importance sampling distribution as
\begin{equation}
q_{\boldsymbol{\eta}}(z_{1:n}|x_{1:n}) = p_{\boldsymbol{\eta}(1)}(z_1|x_1)
\prod_{t=2}^{n}p_{\boldsymbol{\eta}(t)}
\left( z_t | z_{1:t-1}; x_{t} \right).
\end{equation}
After some cancellation, we arrive at
\[
\frac{\boldsymbol{\phi}_{\boldsymbol{\eta}}\left( z_{1:n}\right)}{q(z_{1:n})} = 
\left[
\Phi\left( b_1 \right) -  \Phi\left( a_1 \right) \right] \prod_{t=2}^{n}\left[  \Phi\left( \dfrac{b_t-\hat{z}_t }{r_t}  \right)    - \Phi\left(\dfrac{a_t-\hat{z}_t }{r_t} \right)  \right].
\]
Here, $\phi_{\boldsymbol{\theta}}(z_{1:n})$ denotes the multivariate normal distribution with a zero mean and covariance matrix that of $Z_{1:n}$.  See \cite{kong2023seasonal} for derivation details.

Define the initial weight $w_1 = \Phi(b_1) - \Phi(a_1)$.  The weights are recursively updated via 
\[
w_t = w_{t-1}
\left[
\Phi\left( \frac{b_t-\hat{z}_t }{r_t} \right)   - 
\Phi\left( \frac{a_t-\hat{z}_t }{r_t} \right)
\right]
\]
at time $t$ during the sequential sampling procedure. At the end of the sampling, we obtain 
\[
w_n = \frac{\boldsymbol{\phi}_{\boldsymbol{\eta}}( z_{1:n})}{q_{\boldsymbol{\eta}}(z_{1:n}|x_{1:n})}.
\]
In the classic GHK simulator, $\hat{Z}_t$ and $r_t$ are obtained from the covariance matrix of $\{ Z_t \}$. When $\{ Z_t \}$ is a causal autoregression of order $r$, viz., 
\[
Z_t = \phi_1 Z_{t-1} + \dots + \phi_p Z_{t-r} + \epsilon_t
\]
where $\{ \epsilon_t \}$ is Gaussian white noise with a variance $\sigma^2_\epsilon$ that induces $\mbox{Var}(Z_t)=1$, the one-step-ahead predictors and their mean squared errors obey
\[
\hat{Z}_t=\phi_1 Z_{t-1} + \dots + \phi_r Z_{t-r}, \quad t > r,
\]
and $r_t = \sigma_\epsilon$ for $t > r$.   See 
\cite{Brockwell_Davis_1991} for computing these quantities when $t \leq r$.  

The above procedure generates a fair draw of a single ``particle path" $\{ Z_t \}$ with the property that $\{ X_t \}_{t=1}^n$ generated from $\{ Z_t \}_{t=1}^n$ yields the observations $x_1, \ldots, x_n$. Repeating this process $m$ independent times gives $m$ simulated process trajectories.  Let $\{ {\bf Z}^{(1)}, \ldots, {\bf Z}^{(m)} \}$ denote these trajectories and denote their corresponding time $n$ weights by $\{ w_n^{(k)} \}_{k=1}^m$.

The importance sampling estimate of the likelihood is given by
\[
\hat{\mathcal{L}} \left( \boldsymbol{\theta},\boldsymbol{\eta} \right)=\frac{1}{m}\sum_{k=1}^{m}w_{n}^{(k)}.
\] 
A large $m$ of course provides more accurate estimation. The popular ``BGSF" gradient step and search method is used to optimize the estimated likelihood $\hat{\mathcal{L}}(\boldsymbol{\theta}, \boldsymbol{\eta})$; other optimizers may also work.  

Common random numbers (CRNs), techniques that use the same random quantities across differing parameter values in particle filtering, are used to produce a ``smooth" estimated likelihood function.  With CRNs, Hessian-based standard errors derived from the likelihood function's derivatives at the likelihood estimate are much more reliable; see \cite{kleinman1999simulation} and \cite{glasserman1992some} for more on CRNs.

\subsubsection{A Simulation Study}

This section studies parameter estimators of the superpositioned and Gaussian copula Poisson count series through simulation.  To illustrate the techniques in a simple setting, our simulations consider a single trend and covariate:
\[
\lambda_t = \exp\{ \mu + \beta_1 t + \beta_2 C_t \}.
\]
More complicated scenarios are dealt with similarly. Here, $C_t$ is the value of the covariate at time $t$, generated here as zero-one IID Bernoulli($0.3$) draws under the R seed ``1234".  The covariate sequence $\{ C_t \}_{t=1}^n$ is fixed through all simulations below.  A log link has been used to keep the Poisson parameter non-negative, with $e^\mu$ being the baseline value of $\lambda$. The quantity $\beta_1$ is the ``trend" parameter and $\beta_2$ measures the contribution of the covariate to the mean.  In general, we do not look to conduct inferences about the location parameter $\mu$. In practice, $\lambda_t$ can be any non-negative function, making the model flexible.

In our simulations below, we set the parameters to $\mu=1$, $\beta_1=0.01$, and $\beta_2=1$ and consider series lengths of 50, 100, and 300.  Five hundred independent simulation replicates are studied in every simulation scenario.

For the superposition scheme, the $\{ B_t \}$ process used to generate our series is obtained from a clipped AR(1) $\{ Z_t \}$ series.  The AR(1) parameters are set to $\phi=1/2$ and $\sigma^2=3/4$ so that a unit variance Gaussian $\{ Z_t \}$ series is clipped.  Here $p$ is forced to 1/2 by setting $B_t = 1_{(0,\infty)}(Z_t)$. During estimation, the autocovariance parameters are fixed to their true values in the linear prediction scheme and we examine estimates of the three parameters ($\mu, \beta_1$, and $\beta_2$) appearing in the mean $\lambda_t$.  

\begin{figure}[h!]
\centering
\includegraphics[width=10cm]{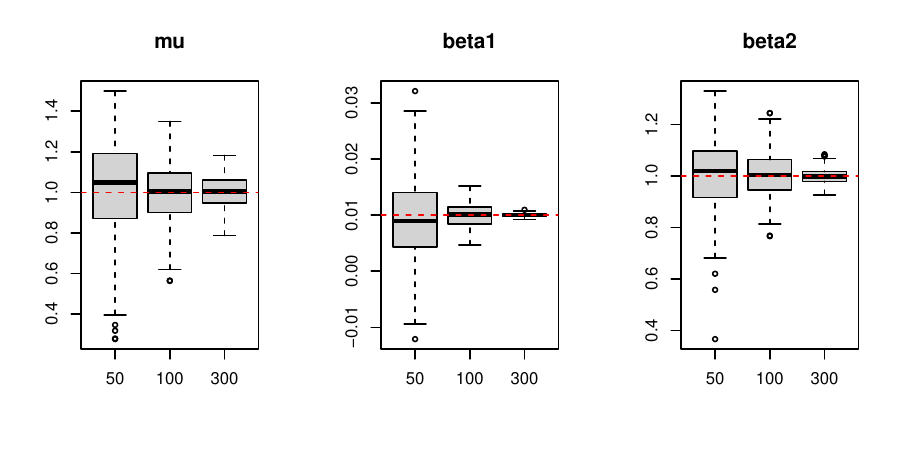}
\caption{Boxplots of parameter estimators for a superpositioned Poisson count series with $\{ B_t \}$ constructed by clipping an AR(1) $\{ Z_t \}$. Both $\phi$ and $p$ are set to 1/2, their true values, during estimation. The three mean parameter estimators appear approximately unbiased; dashed lines demarcate true parameter values.}
\label{fig:Superposition_Poisson_box_plot}
\end{figure}

Figure \ref{fig:Superposition_Poisson_box_plot} displays parameter estimator boxplots for each mean parameter. The dotted red line demarcates the true parameter value. All boxplots are centered around their true parameter values and the distributional shape seems approximately normal. For standard errors of these estimators, Table \ref{tab:Superposition_PoissonAR(1)} reports the sample standard deviations of the parameter estimators over the five hundred runs. As expected, estimation accuracy increases as the series length increases. Overall, the estimators seem accurate.

\begin{table}[hbt!]
\centering
\begin{tabular}{|c|c|c|c|c|}
			\headrow
			\hline
			\multicolumn{5}{|c|}{Superposition Poisson AR(1) Model} \\
			n&&$ \hat{\mu} $&$ \hat{\beta_1} $&$ \hat{\beta_2} $\\ 
			50&mean&1.00805& 0.00914& 1.00887 \\ 
			&SD&0.24117& 0.00733& 0.14888 \\ \hline
            100&mean&0.99380 &0.01003& 1.00240 \\ 
			&SD&0.14921& 0.00210& 0.08668 \\ \hline
            300&mean&1.00271& 0.00999& 1.00012 \\ 
			&SD&0.08064& 0.00031& 0.03107 \\ \hline
		\end{tabular}
\caption{Mean and standard deviation (SD) of estimators over 500 independent runs for the superpositioned Poisson count series with $\{ B_t \}$ constructed via a clipped AR(1). True values of the parameters are $\mu=1$, $\beta_1=0.01$, and $\beta_2=1$. The results report the sample mean and standard deviation (denominator of $499$) of the parameter estimates.}
\label{tab:Superposition_PoissonAR(1)}
\end{table}

Moving to the Gaussian copula scheme, we use an AR(1) series with $\phi=1/2$ and $\sigma^2=3/4$ as the latent process $\{ Z_t \}$.  The settings for $\mu, \beta_1$, and $\beta_2$ used above are repeated.  In this scheme, all parameters are estimated via particle filtering methods, even the AR(1) parameter $\phi$.

\begin{figure}[h!]
\centering
\includegraphics[width=10cm]{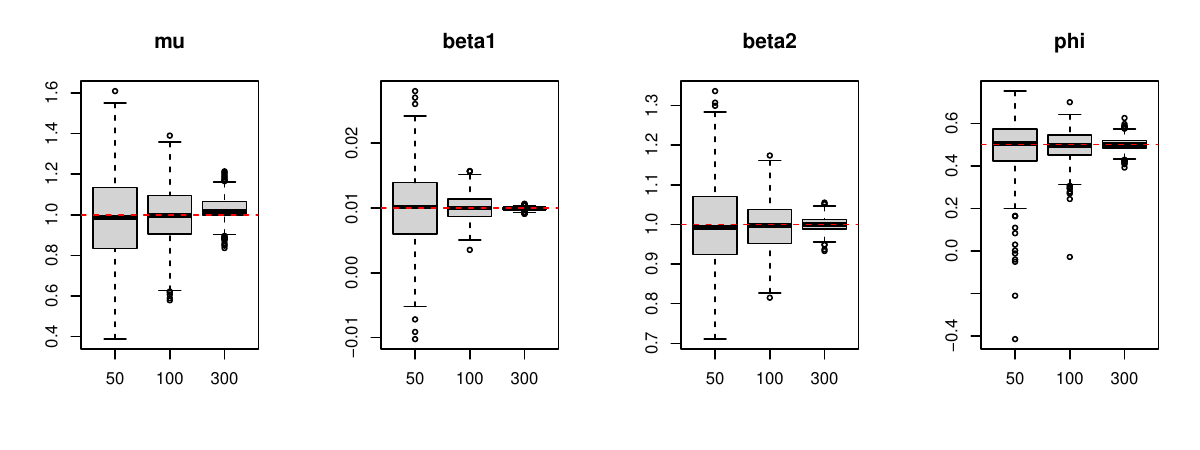}
\caption{Boxplots of parameter estimators for Gaussian copula estimates of a Poisson count series with an AR(1) $\{ Z_t \}$ with $\phi = 0.5$. All estimators appear approximately unbiased; dashed lines demarcate true parameter values.}
\label{fig:Poisson_trend_plot}
\end{figure}

Figure \ref{fig:Poisson_trend_plot} shows boxplots of all estimators and series lengths. The dotted red line again indicates true parameter values. All boxplots are centered around the true parameter values and look approximately normal, with perhaps an exception being $\hat{\phi}$ under the shortest series length $n=50$.  For standard errors, Table \ref{tab:PoissonAR(1)} reports two values: 1) the sample standard deviation of the parameter estimators over the five hundred independent runs (denominator of 499), and 2) the average (over all runs) of standard errors obtained by inverting the Hessian matrix at the maximum likelihood estimate for each run (denominator of 500). The difference between these two values are quite small, implying that Hessian-based standard errors obtained from one sample path are indeed accurate. Standard errors again decrease with increasing $n$.  Again, the performance appears good.

\begin{table}[hbt!]
\centering
\begin{tabular}{|c|c|c|c|c|c|}
			\headrow
			\hline
			\multicolumn{6}{|c|}{Gaussian Copula Poisson AR(1) Model} \\
			n&&$ \hat{\mu} $&$ \hat{\beta_1} $&$ \hat{\beta_2} $&$ \hat{\phi} $\\ 
			&mean&0.98419& 0.01012& 0.99880 &0.48945 \\ 
			50&SD&0.22609& 0.00610& 0.11068& 0.13010 \\ 
			&$ \hat{E}(I'(\theta)^2) $&0.22640 &0.00644& 0.10701& 0.11412\\ \hline
            &mean&0.99550 &0.01004& 0.99716& 0.49222 \\ 
			100&SD&0.13941 &0.00200& 0.06520& 0.07693 \\ 
			&$ \hat{E}(I'(\theta)^2) $&0.14720& 0.00206& 0.06642& 0.07498\\ \hline
            &mean&1.02786& 0.00988& 1.00020& 0.50182 \\ 
			300&SD&0.05987& 0.00025& 0.01896& 0.03320 \\ 
			&$ \hat{E}(I'(\theta)^2) $&0.06410& 0.00027& 0.02219& 0.04144\\ \hline
		\end{tabular}
\caption{Standard errors for the parameter estimators for the Poisson marginal distribution with an AR(1) $\{ Z_t \}$.  The results show the sample standard deviation (SD) of the parameter estimators from five hundred independent series (denominator of 499), and the average of the five hundred  standard errors obtained by inverting the Hessian matrix ($\hat{E}[ I^\prime(\theta)^2)$] at the maximum likelihood estimate over these same runs.}
\label{tab:PoissonAR(1)}
\end{table}

In comparing superpositioned and Gaussian copula results, the standard errors for the Gaussian likelihood estimators are slightly smaller than their superpositioned counterparts.  This is expected:  likelihood estimators are generally the asymptotically most efficient estimators.  This said, the calculations needed to produce the likelihood estimators are more intensive than those for linear prediction.  Finally, we did study higher order autoregressions; results are again impressive and similar to the above. For brevity's sake, figures and tables of these simulations are omitted.

\section{Applications}
\label{App}

This section considers two count series that we fit with Poisson marginal distributions:  Atlantic Basin tropical storm counts and the number of no-hitter games pitched annually in Major League Baseball.   Both series are comprised of small counts, where the marginal distribution becomes important.  Because superpositioned linear prediction estimation performs slightly worse than Gaussian copula likelihood estimation, the latter technique is concentrated upon in this section.

\subsection{Atlantic Tropical Cyclones}

\begin{figure}[h!]
\centering
\includegraphics[width=14cm]{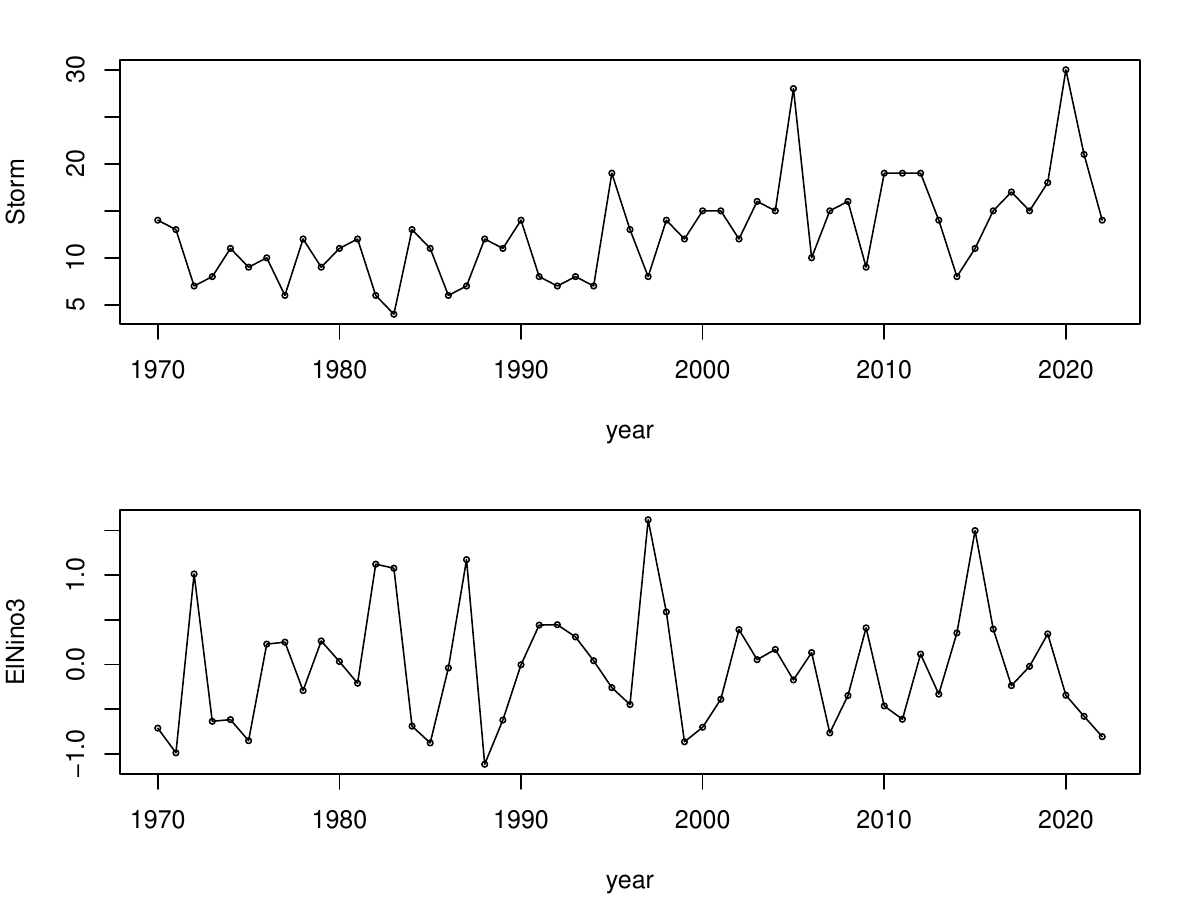}
\caption{Top: the yearly number of North Atlantic Basin tropical storms recorded from 1970-2022. Bottom: the annual El-Nino covariate (ElNino3) over the same period.}
\label{fig:tropical_data}
\end{figure}

Our first series contains the annual number of tropical cyclones observed in the North Atlantic Basin since 1970.  This series is plotted in the top plot of Figure \ref{fig:tropical_data}. Poisson marginal distributions have been previously advocated for these and other tropical cyclone counts \citep{Robbins_etal_2011, solow2008incompleteness, mooley1980severe}.   

There is concern that the number of North Atlantic Basin cyclones has been increasing in recent years, with researchers pointing to 1995 as a year where the North Atlantic warmed and tropical storm activity increased; see the changepoint analyses in \cite{Robbins_etal_2011} and \cite{Fisher_etal_2020}. Because of this, we will allow for a linear trend as one covariate (a changepoint mean shift structure is also worthy of consideration).  A strong El-Nino index, which is a measure of equitorial warming in the Pacific Ocean, is thought to impede Atlantic tropical cyclone development \citep{gray1984atlantic, goldenberg1996physical} through its influence on the southern jet stream:  a strong El-Nino produces a strong southern jet stream, which produces wind shear at stratospheric levels, shearing tops of thunderstorm clouds off and hindering tropical cyclone development.  As a second covariate, annual values of El-Nino 3, which are shown in the bottom plot of Figure \ref{fig:tropical_data}, are used.

While the North Atlantic Basin tropical cyclone record goes back to 1851, some of the earliest data is thought to be incomplete. Specifically, there is concern that some weak storms formed over the central Atlantic waters in the record's earliest years, lived their entire lives over open water, and were never detected. As such, we start our analysis at 1970.  This is approximately when the GOES satellites were launched; no storms should have evaded detection thereafter. While one could use indicator variables as additional covariates to quantify undetected storms, we will simply start the record at 1970; see \cite{Robbins_etal_2011} and \cite{Fisher_etal_2020} for an analysis of the Atlantic Basin's full record.

The level of correlation in this count series is not extreme.  In fact, many authors view the annual counts as approximately independent \citep{Robbins_etal_2011}.  Certainly, if significant year-to-year autocorrelation existed, storm counts would be easier to forecast a year in advance. (Annual forecasting competitions are conducted in May for this series, where Poisson regression methods are typically used with various meteorological covariates to predict counts for the upcoming June-November season.  Forecasts even a year in advance have generally shown little predictive power). Our model fits below will confirm that there is minimal year-to-year autocorrelation in these counts.

\begin{table}[hbt!]
\centering
\begin{tabular}{|c|c|c|c|c|c|c|c|c|}
			\headrow
			\hline
			\multicolumn{9}{|c|}{Model: $\lambda_t = e^{\mu +\beta_1t + \beta_2C_t}$} \\
			&& $\hat{\mu}$ &$\hat{\beta}_1$ &$\hat{\beta}_2$ & $\hat{\phi}_1$ &$\hat{\phi}_2$ & AIC & BIC \\ 
			WN&Est.&2.0699 & 0.0154& -0.2830 &NA&NA&{\bf 283.4695}&{\bf 289.3803} \\ 
			&$\sqrt{E(I'(\hat{\theta})^2)}$&0.0872& 0.0025 &0.0658& NA&NA&&\\ \hline
            AR1&Est.&2.0699 & 0.0154& -0.2831& -0.0018 &NA&285.4694&293.3506\\ 
			&$\sqrt{E(I'(\hat{\theta})^2)}$&0.0874& 0.0025& 0.0660 &0.1623 &NA&&\\  \hline
            AR2&Est.&2.0598&  0.0158& -0.2839& -0.0456& -0.2326&286.6195&296.471\\ 
			&$\sqrt{E(I'(\hat{\theta})^2)}$&0.0760 &0.0022 &0.0673 &0.1628 &0.1553&&\\ \hline
		\end{tabular}
\caption{}
\label{tab:Tropical_model}
\end{table}

The results are as expected. First, there is little autocorrelation in these counts.  Here, we fitted white noise, AR(1), and AR(2) autocorrelation structures in the latent Gaussian process, but both AIC and BIC model selection criteria in Table \ref{tab:Tropical_model} prefer the white noise model. With this white noise structure, the estimated trend parameter in the model is $\hat{\beta}_1 = 0.0154~(0.0025)$, which translates to a hurricane season that will be some four and a half times more active in 2070 than it was in 1970.  The standard error of this estimator produces a $z$-score of about 6.2, indicating a significant increasing trend in the counts and trouble for coastal residents.  The estimated coefficient of the El-Nino covariate is $\hat{\beta}_2=-0.2830$ with a standard error of 0.0658.  This parameter is significantly negative, with a $z$-score of about -4.3.  Indeed, an active El-Nino appears to impede Atlantic tropical cyclone development. 

\subsection{Baseball No-hitters}

Our second series contains the number of annual no-hitter games pitched in major league baseball from 1893 - 2022.   A no-hitter occurs when a pitcher (or multiple pitchers) do not allow the opposing team to get any hits over the course of a game.  It is indicative of a dominant pitching performance.

\begin{figure}[h!]
\centering
\includegraphics[width=14cm]{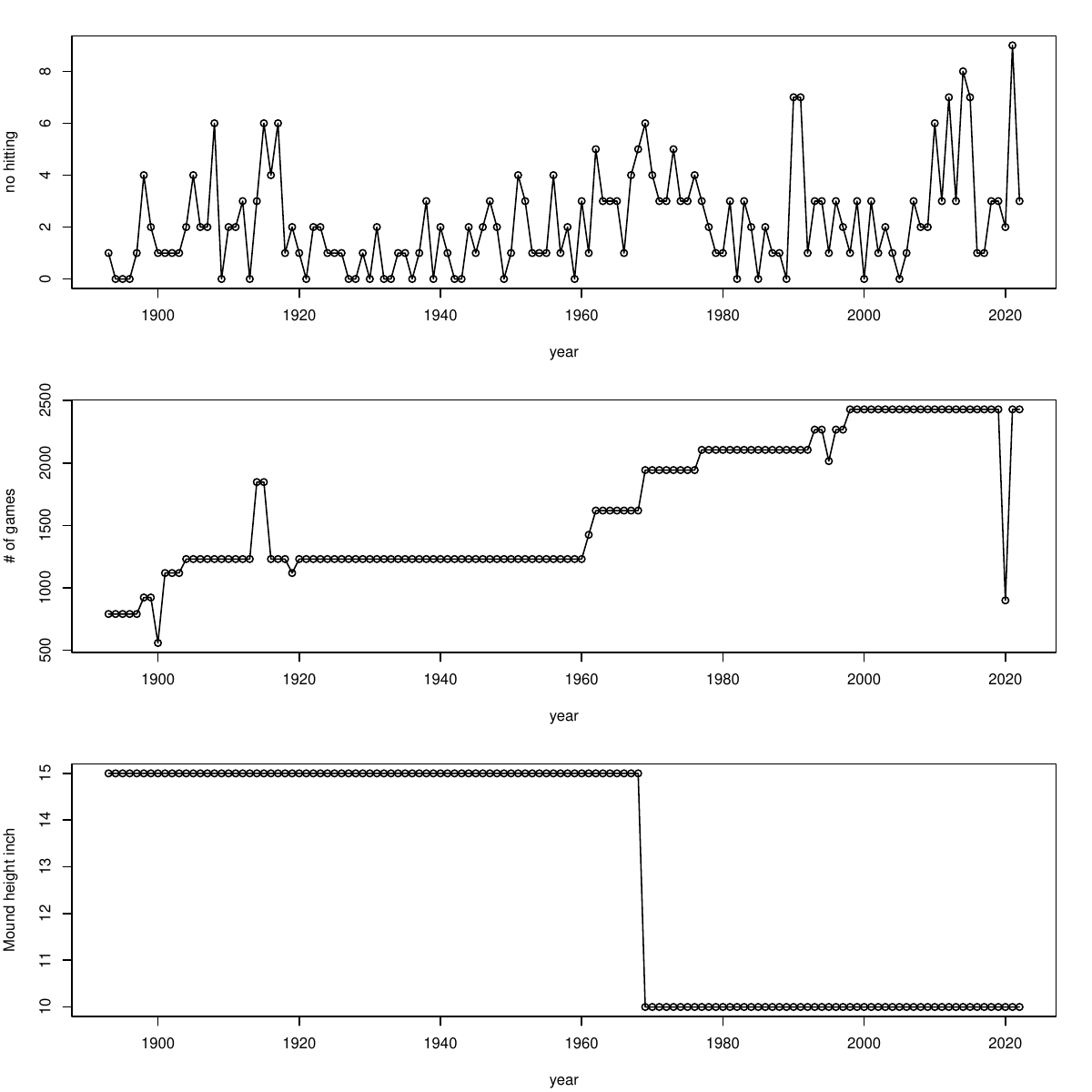}
\caption{Top: Major League baseball annual no hitter counts from 1893-2022. Middle: The number of games played during each year by all teams.  Bottom: The height of the pitching mound in inches during each year.}
\label{fig:no_hitter_data}
\end{figure}

There has never been more than nine no-hitters pitched in a season; some years do not see any non-hitters.  Figure \ref{fig:no_hitter_data} shows the no-hitter counts along with two explanatory covariates:  the total number of games played in the major league baseball season and the height of the pitching mound.  The total number of games played in has changed by season as more teams have been added to the league; also the number of games that teams play in a season has varied.  Strikes and the Covid-19 pandemic have forced cancellation of some games in a few sporadic years.   Of course, the more games played, the more likely it is to have a no-hitter pitched. Our second covariate is the height of the mound. A higher pitching mound is thought to give pitchers an advantage. The height of the pitching mound was reduced from 15 inches to 10 inches in 1969; hence, this covariate could be viewed as a breakpoint or intervention (known changepoint).

Our model here takes 
\[
\lambda_t = \exp\{   \mu + \beta_1 C_{1,t} + \beta_2 C_{2,t} \},
\]
where $C_{1,t}$ is the number of games played in year $t$ and $C_{2,t}$ is the height of the pitching mound in year $t$. As we will see, there is some autocorrelation in these counts.

Table \ref{tab:Nohitter_1} shows the results of the Gaussian copula model fit with white noise, AR(1), and AR(2) errors for $\{ Z_t \}$.  First, both AIC and BIC model selection statistics prefer an AR(1) $\{ Z_t \}$.  The estimated AR(1) coefficient here is $\hat{\phi}=0.3199$, which is more autocorrelation than we perhaps expected (no-hitters are extreme performances and rare, which are often modeled as independent; see the peaks over threshold theory in \cite{pickands1975statistical}).  While we do not consider eliminating the mean $\mu$ in the model, the estimates and standard errors for $\beta_2$ suggest that pitching mound height does not significantly influence no-hitter counts, but that more no-hitters occur when more games are played.    

\begin{table}[hbt!]
\centering
\begin{tabular}{|c|c|c|c|c|c|c|c|c|}
			\headrow
			\hline
			\multicolumn{9}{|c|}{Model: $\lambda_t = e^{\mu+\beta_1C_{1,t}+\beta_2C_{2,t}}$} \\
			&&$ \hat{\mu} $&$ \hat{\beta_1} $&$ \hat{\beta_2} $&$ \hat{\phi}_1 $&$ \hat{\phi}_2 $&AIC&BIC\\ 
			WN&Est.&-1.5367 & 0.0008374& 0.0697988 &NA&NA&491.6805&500.2831 \\ 
			&$\sqrt{E(I'(\hat{\theta})^2)}$&1.1982 & 0.0002743 &0.0590602& NA&NA&&\\ \hline
            AR1&Est.&-1.7639 & 0.0008968& 0.0803000 & 0.3199 &NA&486.4131&497.8832\\ 
			&$\sqrt{E(I'(\hat{\theta})^2)}$&1.3005 & 0.0002958 &0.0646546&  0.0710 &NA&&\\  \hline
            AR2&Est.& -1.6198 & 0.0008709& 0.0726532&  0.2792&  0.1531&486.7636&501.1013\\ 
			&$\sqrt{E(I'(\hat{\theta})^2)}$&1.3288&  0.0003008 &0.0664830 & 0.0719&  0.0750&&\\ \hline
		\end{tabular}
\caption{Summary of the No-hitter Poisson count fit. The AIC and BIC model selection criteria prefer AR(1) errors; the pitching mound height covariate appears insignificant.}
\label{tab:Nohitter_1}
\end{table}

Table \ref{tab:Nohitter_2} refits the model with the no-hitter covariate eliminated and AR(1) errors.  The estimators, standard errors, and conclusions do not change appreciably from the last table.  

\begin{table}[hbt!]
\centering
\begin{tabular}{|c|c|c|c|c|c|c|}
			\headrow
			\hline
			\multicolumn{7}{|c|}{Model: $\lambda_t = e^{\mu+\beta_1C_{1,t}}$} \\
			&&$ \hat{\mu} $&$ \hat{\beta_1} $&$ \hat{\phi}_1 $&AIC&BIC\\ \hline
            AR1&Est.&-0.1851 & 0.0005687 & 0.3152 &{\bf 485.9015}&{\bf 494.5041}\\ 
			&$\sqrt{E(I'(\hat{\theta})^2)}$&0.2419& 0.0001284 &0.0706 &&\\  \hline
		\end{tabular}
\caption{A refit of the model in the last table with the pitching mound height covariate eliminated.}
\label{tab:Nohitter_2}
\end{table}

\section{Diagnostics}
\label{Diagnostics}

One issue has been left hanging in our development. This section shows how to test whether or not the Poisson marginal distribution is adequate. The data fits of the last two sections will be revisited. We will concentrate on the Gaussian copula model since that is the most flexible.

\begin{figure}[h!]
\centering
\includegraphics[width=14cm]{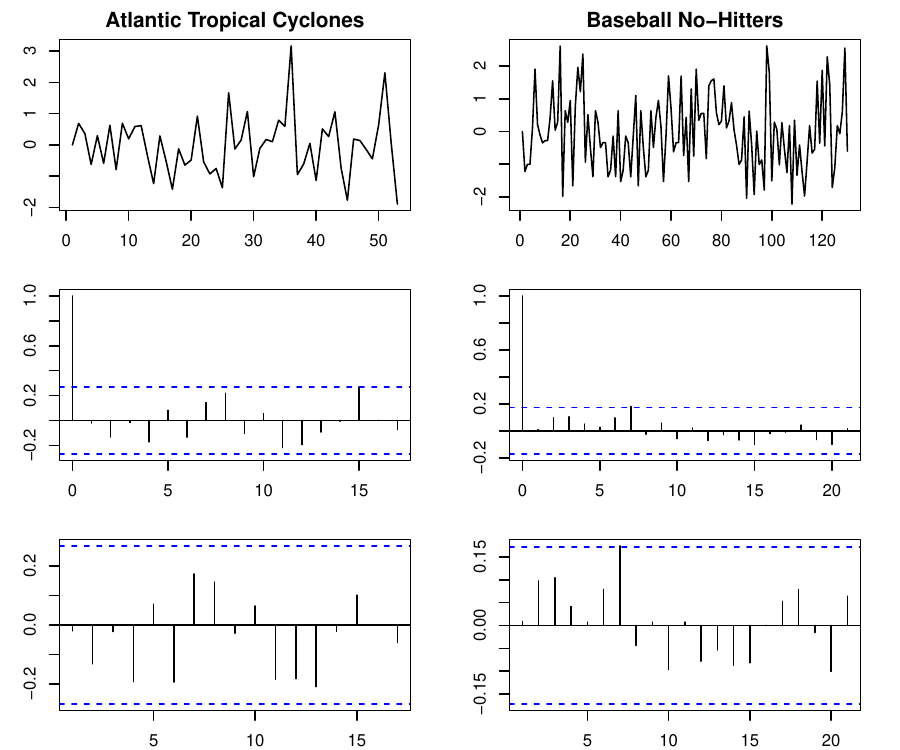}
\caption{Top: Residuals for the Atlantic Basin tropical cyclone (left) and no-hitter (right) counts. Middle: Residual autocorrelations.  Bottom: Residual partial autocorrelations.}
\label{fig:res}
\end{figure}

A simple definition of a model residual tries to recover the latent $\{ Z_t \}$ process and mimic autoregressive residuals.  The conditional expectation 
\begin{equation}
\label{residual}
\mathbb{E} [ Z_t | X_t ] =
\frac{\exp(-\boldsymbol{\Phi}^{-1}(F_{\lambda_t}^2(x_t-1)/2)-\exp(-\boldsymbol{\Phi}^{-1}(F_{\lambda_t}^2(x_t)/2)}{\sqrt{2\pi}(F_{\lambda_t}(x_t)-F_{\lambda_t}(x_t-1))} := \hat{Z}_t
\end{equation}
is an estimate of $Z_t$ from $X_t$ only.  While a better residual would use $E[Z_t|X_1, \dots, X_t]$, this quantity appears intractable and this definition will prove sufficient for our purposes.  The fitted autoregressive model and $\{ \hat{Z}_t \}$ can be used to define the residuals. These are simply
\[
\hat{R}_t := \hat{Z}_t - \sum_{k=1}^r \hat{\phi}_k \hat{Z}_{t-k}, \quad t > r. 
\]
Figure \ref{fig:res} plots these residuals for our best fitting models for the baseball and tropical cyclone series, along with sample correlations and partial autocorrelations.  Point-wise ninety five percent confidence bands for white noise are included in the plot. No autocorrelation is appreciably evident in these residuals.

To assess adequacy of the Poisson marginal distribution, probability integral transforms (PIT) techniques can be be used.  PIT methods were proposed in \cite{dawid1984present} and assess the consistency between probabilistic forecasts of the individual observations from the fitted model and the observations themselves.  When the predictive distribution is continuous, PIT residuals are uniformly distributed over $(0,1)$. We will use the nonrandomized PIT residuals in \cite{czado2009predictive}, where uniformity still holds in the discrete case. 

PIT residuals begin with the conditional cumulative distribution function of $X_t$:
\begin{equation}
P_t(y):=\mathbb{P} 
\left( X_t \leq y | 
X_1 = x_1, \dots, X_{t-1}=x_{t-1} \right), \quad
y \in \left\{ 0, 1, \ldots \right \}.
\end{equation}
Then the nonrandomized mean PIT residual is the sample average
\[
\bar{F}(u)= n^{-1}\sum_{t=1}^{n}F_t(u|x_t),
\]
where
\[
F_t(u|y)=\left\{\begin{array}{cl}
0,&\hbox{if }u\leq P_t(y-1)\\
\dfrac{u-P_t(y-1)}{P_t(y)-P_t(y-1)},&\hbox{if }P_t(y-1)<u<P_t(y)\\
1,&\hbox{if }u\geq P_t(y)
\end{array}
\right..
\]
The quantity $P_t(y)$ can be approximated during the particle filtering likelihood evaluation algorithms; specifically,
\[
\hat{P}_t(y) = \sum_{i=0}^{y}w_{i,t}(\hat{Z}_t),
\]
where 
\[
w_{i,t}(z)=
\Phi\left(\dfrac{\Phi^{-1}(F_{\lambda_t}(i))-z}{r_t} \right)-
\Phi\left(\dfrac{\Phi^{-1}(F_{\lambda_t}(i-1))-z}{r_t} \right).
\]
The weight $w_{i,t}(z)$ can be obtained at time $t$ from the particle filtering algorithm. 
\begin{figure}[h!]
\centering
\includegraphics[width=12cm]{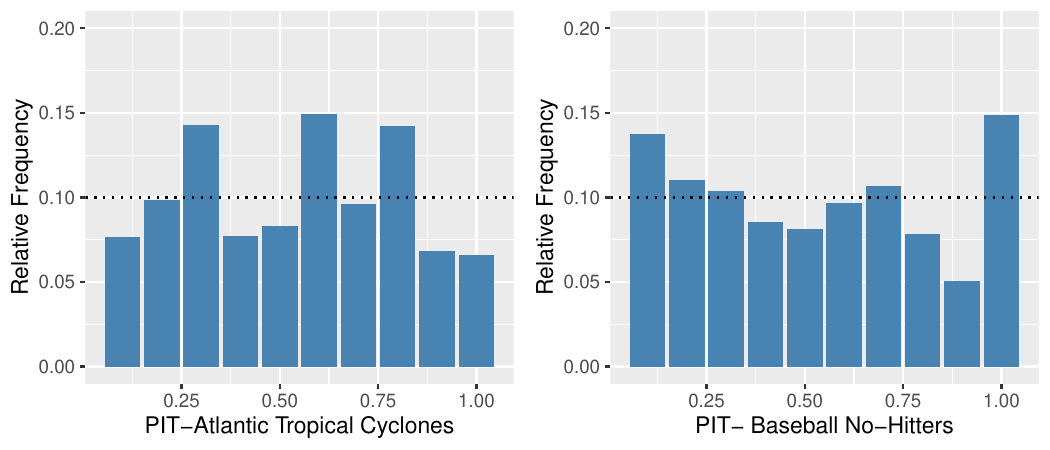}
\caption{PIT plot for the best fitted models. Left: PIT histogram for the Atlantic Basin tropical cyclones series with $\lambda_t = \exp\{ \mu+\beta_1C_{1,t}+\beta_2C_{2,t} \}$ and white noise errors; Right: PIT histogram for the baseball no-hitter series with $\lambda_t = \exp \{ \mu+\beta_1 C_{1,t} \}$ and AR(1) errors.}
\label{fig:diagonostic}
\end{figure}

To assess this fit, we report a $p$-value based on of a deviation from a uniform histogram over [0,1] containing 10 equally spaced bins.  Our statistic is
\[
Q:=\frac{1}{10} \sum |\hat{f}_i - 1/10|,
\]
where $\hat{f}_i$ is the proportion of residuals falling in the the category $((i-1)/10, i/10]$ for $i=1, \dots, 10.$  The statistics are $Q=0.0270$ for the Atlantic cyclone data and $Q=0.0216$ for the no-hitter data. A Poisson marginal distribution is rejected when $Q$ is too large.  Our $p$-values, which were computed via simulation, are $0.76$ for the Atlantic cyclone data and $0.37$ for the no-hitter data. One sees little reason to doubt a Poisson marginal distribution with either series.  When a PIT residual analysis rejects a Poisson marginal, it is not clear to us how to modify the marginal distribution from the PIT plot; however, the Gaussian copula techniques here apply to non-Poisson count distributions.

\section{Concluding Comments}
\label{Comments}

This paper studied some methods that produce time series of Poisson distributed counts. Both stationary and non-stationary settings were considered and inference methods for some of the well-performing model classes were developed, including testing the Poisson marginal assumption.  Many of the classical methods have deficiencies in what they can handle.  An implication of the paper is that the Gaussian copula transformation technique is the most flexible paradigm considered as it produces the most general autocovariance structures possible, easily accommodates covariates, and likelihood methods of inference can be conducted via particle filtering methods.  The popular INAR model class was deemed deficient in several manners.

Additional research is needed on several fronts.  First, ways to generate Poisson counts beyond those discussed here exist.  Worthy of mention are stationary Markov chain techniques \citep{Xiaotian} and shot noise methods \citep{jang2021review}, the latter being related to our superpositioning techniques here.  Given the flexibility of the Gaussian copula paradigm, it may be pedantic to investigate these classes further unless they can be shown to be flexible, parsimonious, accommodate covariates, and have analyzable likelihood functions.  Second, asymptotic normality of the parameter estimators was not proven here, but needs to investigated. We are unsure how to do this when the likelihood function is intractable as in the Gaussian copula setting.  Third, multivariate versions of the methods are worthy of development.  Here, one needs to settle on a definition of multivariate Poisson --- many are possible \citep{teicher1954multivariate, kocherlakota2017bivariate, inouye2017review}.   Finally, extensions of the methods to the zero inflated case, which frequently arises with Poisson analyses \citep{lambert1992zero, fernando2022review}, are worth considering.

\section*{Acknowledgements} The authors thank NSF Grant DMS 2113592 for partial support of this research.

\section{Data Availability Statement}
All data series and code are available from the authors upon request.

\bibliography{References}

\end{document}